\documentclass[12pt]{article}
\usepackage{graphicx}
\usepackage[latin1]{inputenc}
\usepackage{amsfonts}
\usepackage{epsfig}
\usepackage{amsmath}

\advance\textwidth by 3truecm
\advance\oddsidemargin by -1.5truecm

\begin{document}

\title{\normalsize\bf LETTER\\[\baselineskip]
Statistical correlations of an anyon liquid at low temperatures}
\author{\normalsize F. M. D. Pellegrino$^1$,
G. G. N. Angilella$^2$\footnote{Corresponding author.
E-mail: {\tt giuseppe.angilella@ct.infn.it}}, N. H. March$^{3,4}$, R. Pucci$^2$\\
\normalsize\emph{$^1$Scuola Superiore di Catania, Universit\`a di
   Catania,}\\
\normalsize\emph{Via S. Nullo, 5/i, I-95123 Catania, Italy,}\\
\normalsize\emph{and INFN, Sez. Catania, Via S. Sofia, 64, I-95123 Catania, Italy}\\
\normalsize\emph{$^2$Dipartimento di Fisica e Astronomia, Universit\`a di
   Catania,}\\
\normalsize\emph{and CNISM, UdR Catania, and INFN, Sez. Catania,}\\
\normalsize\emph{Via S. Sofia, 64, I-95123 Catania, Italy}\\
\normalsize\emph{$^3$Oxford University, Oxford, UK}\\
\normalsize\emph{$^4$Department of Physics, University of Antwerp,}\\
\normalsize\emph{Groenenborgerlaan 171, B-2020 Antwerp, Belgium}}

\date{\normalsize (Received on 8 February 2008, in final form)}

\maketitle

\abstract{%
Using a proposed generalization of the pair distribution function for a gas of
non-interacting particles obeying fractional exclusion statistics in arbitrary
dimensionality, we derive the statistical correlations in the asymptotic limit
of vanishing or low temperature. While Friedel-like oscillations are present in
nearly all non-bosonic cases at $T=0$, they are characterized by exponential
damping at low temperature. We discuss the dependence of these features on
dimensionality and on the value of the statistical parameter
$\alpha$.\\[0.5\baselineskip]
{\sl Keywords:} Electron liquid; Quantum Hall effect; Fractional statistics;
Pair correlation function.\\[0.5\baselineskip]
{\sl Running head:} {\bf Low-temperature correlations of anyon liquids.}
}

\newpage

Fractional statistics has been introduced to describe the fractional quantum
Hall effect in low-dimensional, correlated electron quantum liquids.
\cite{Leinaas:77,Wilczek:82,Wilczek:82a,Laughlin:83,Laughlin:83a,Laughlin:88}
The elementary excitations of such systems, usually dubbed anyons, can be
characterized either through the phase factor $e^{i\alpha\pi}$ ($0\leq \alpha
\leq1$) picked up by the overall wave-function upon exchange of two particles,
of by the variation of size $\Delta D$ of the available Hilbert space
corresponding to a variation $\Delta N$ of the number of particles, through the
ratio $g=-\Delta D/\Delta N$ ($0\leq g \leq1$). The limiting cases $\alpha=0$,
$g=0$ [$\alpha=1$, $g=1$] correspond to boson [fermion] statistics,
respectively. While the former characterization, corresponding to
\emph{exchange} fractional statistics, applies to dimensions $d \leq 2$ only,
the latter characterization, corresponding to \emph{exclusion} statistics, does
not suffer from such a restriction, and has been considered for arbitrary
dimensionality. The relation between the two definitions of fractional
statistics is not completely settled, to date \cite{Speliotopoulos:97}. 

In this context, it has been recently proposed that topological excitations such
as vortex rings (anyonic loops) in the three-dimensional chiral spin liquids may
obey fractional non-Abelian statistics \cite{Si:07}. For the sake of simplicity,
and in order to avoid confusion with the pair correlation function to be defined
below, in the following we shall use the symbol $\alpha$ for the statistical
parameter in both contexts.

Recently, the statistical correlations of an ideal gas of particles obeying
fractional exchange statistics have been addressed, in arbitrary dimensionality,
as a function of density, temperature, and statistical parameter $\alpha$,
partly analytically and partly numerically, through the study of the pair
correlation function $g(\textbf{r})$ \cite{Pellegrino:07}. This function
represents the probability of simultaneously finding two anyons at positions
$\textbf{r}$ and $\textbf{r}=0$, respectively. A general result, applying to all
non-bosonic values of the fractional parameter ($0 < \alpha \leq1$), is that the
pair correlation function is characterized by Friedel-like oscillations, which
become more pronounced for increasing $\alpha$ or density, or for decreasing
$T$. A somewhat related study \cite{Sen:07} considered the two-particle kernel
in the lowest Landau level of a quantum Hall system, but now in $d=2$ and within
the context of exchange fractional statistics.

In this Letter, we study some asymptotic properties of the pair correlation
function at zero temperature as a function of density.

In the formalism of second quantization, the pair correlation function introduced above can be defined as:
\begin{equation}
g(\textbf{r}) = \frac{\langle \Psi^\dag (\textbf{r}) \Psi^\dag (0) \Psi (0)
\Psi(\textbf{r})\rangle}{n(\textbf{r}) n(0)},
\label{eq:gr}
\end{equation}
where $\Psi^\dag (\textbf{r})$ [$\Psi(\textbf{r})$] is a creation [annihilation] quantum field operator at position $\textbf{r}$, $n(\textbf{r})=\langle \Psi^\dag(\textbf{r}) \Psi(\textbf{r})\rangle$ is the single-particle probability density  at position $\textbf{r}$, and $\langle ...\rangle$ denotes a quantum statistical average associated with the equilibrium distribution of the identical particle assembly under analysis.

In Ref.~\cite{Pellegrino:07}, we introduce a particular generalization of the
commutation relations for the $\Psi$-fields, interpolating between boson and
fermion statistics for $0\leq\alpha\leq1$ and arbitrary dimensionality $d$.
These are reminiscent of the graded commutation relations of Greenberg
\cite{Greenberg:91}, weakly violating conventional quantum statistics.

In this way, the pair correlation function for a translationally system of
anyons obeying exclusion statistics can be written as:

\begin{eqnarray}
g (r ) &=& 1+\cos(\alpha\pi) \frac{|\langle \Psi^\dag (r )  \Psi (0 ) \rangle|^2}{n^2 (0)} \nonumber\\
&=&  1+\cos(\alpha\pi) \left| \frac{\tilde{n} (r)}{\tilde{n} (0)} \right|^2 ,
\label{eq:galpha}
\end{eqnarray}
where
\begin{equation}
\tilde{n}(r) = \int \frac{d^d \textbf{k}}{(2\pi)^d} e^{-i\textbf{k}\cdot \textbf{r}} n(\textbf{k})
\label{eq:gnFourier1}
\end{equation}
is the Fourier trasform in $d$ dimensions of the distribution function
$n(\textbf{k})$ for exclusion anyons at equilibrium. While alternative
derivations of the latter have been proposed by Ouvry \emph{et al.}
\cite{DasnieresdeVeigy:94,DasnieresdeVeigy:95}, as well as viable approximations
within the chemical collision model by March \emph{et al.}
\cite{March:93c,March:93b,March:97b}, in the following we shall use Wu's
definition \cite{Wu:95}:
\begin{equation} 
n(\textbf{k})=\frac{1}{w(\zeta)+\alpha},
\label{eq:distrib_wu}
\end{equation}
where $\zeta=\zeta(\textbf{k})$, and $w(\zeta)$ satisfies the functional relation:
\begin{equation}
w^\alpha (1+w)^{1-\alpha}=\zeta.
\end{equation}

In this Letter we treat the case of an ideal gas of anyons obeying exclusion exclusion statistics, and for their dispersion relation we assume the  usual one as for a non-relativistic free particle:
\begin{equation} \label{eq:disp_rel}
\epsilon(\textbf{k}) = \frac  {\hbar^2k^2}{2m}.  
\end{equation}

At $T=0$ and for each non-zero value of the statistical parameter $\alpha$
($0<\alpha \leq1$), Wu's distribution function Eq.~(\ref{eq:distrib_wu}) becomes
a step function, as in Fermi statistics. In particular, it assumes the following
limiting form:
\begin{equation} \label{eqn:scalinoanioni}
n(\epsilon)=
\left \{
\begin{array}{rl}
1/\alpha, &  	\epsilon \leq \epsilon_{f,\alpha}, \\
0, & \epsilon > \epsilon_{f,\alpha},
\end{array}
\right.
\end{equation}
where $\epsilon_{f,\alpha}$ is a generalized Fermi energy, that can be defined in arbitrary dimension as the highest level occupied by $N$ particles at $T=0$.
It is appropriate at this point to remind that under the same conditions in the bosonic case ($\alpha=0$) the distribution function behaves like a Dirac delta function, $n(\epsilon)=N\delta(\epsilon)$.

Therefore, inserting Eq.~(\ref{eqn:scalinoanioni}) into
Eq.~(\ref{eq:gnFourier1}) allows us to derive the analytical expression of the
anyon pair correlation function in arbitrary dimension $d$ at $T=0$ as:
\begin{equation} \label{eq:g_zero}
g(r) = 1+\cos(\alpha \pi) \Gamma\left(\frac{d}{2}\right) \left( \frac{2}{k_{f,\alpha} r}\right)^d J^2_{d/2}(k_{f,\alpha} r),
\end{equation}
where $J_{d/2}(z)$ is a Bessel function of first kind of order $d/2$ \cite{GR}. $k_{f,\alpha}$ represents the wavevector implicitly defined by:
\begin{equation} 
C_d \left(\frac{k_{f,\alpha}}{2\pi}\right)^d =\alpha \frac{N}{V},
\label{eq:Wu_Kfermi}
\end{equation}
where $C_d =\frac{\pi^{d/2}}{\Gamma(d/2)}$ denotes the volume of
the unit sphere in $d$ dimensions.
In the fermion case ($\alpha=1$), $\hbar k_{f,\alpha}$ represents the Fermi momentum.

In order to compare the cases with different values of the statistical parameter $\alpha$, we write the pair correlation function as:
\begin{equation} \label{eq:g_zero2}
g(y) = 1+\cos(\alpha \pi) \Gamma\left(\frac{d}{2}\right)   \frac{2^d}{\alpha y^d}   J^2_{d/2}(\alpha^{1/d}y),
\end{equation}
where $y=k_f r$ is a dimensionless variable, with $k_f$ the Fermi wavevector,
which agrees with Eq.~(\ref{eq:Wu_Kfermi}) when $\alpha=1$. For $0<\alpha\leq1$,
$g(y)$ has the same functional form of the fermionic limit, the differences
being the factor $\cos(\alpha \pi)/\alpha$ and the dependence of $k_{f,\alpha}$
on the statistical parameter $\alpha$.

\begin{figure}[t]
\begin{center}
\begin{minipage}[c]{0.49\textwidth}
\begin{center}
\includegraphics[width=\textwidth]{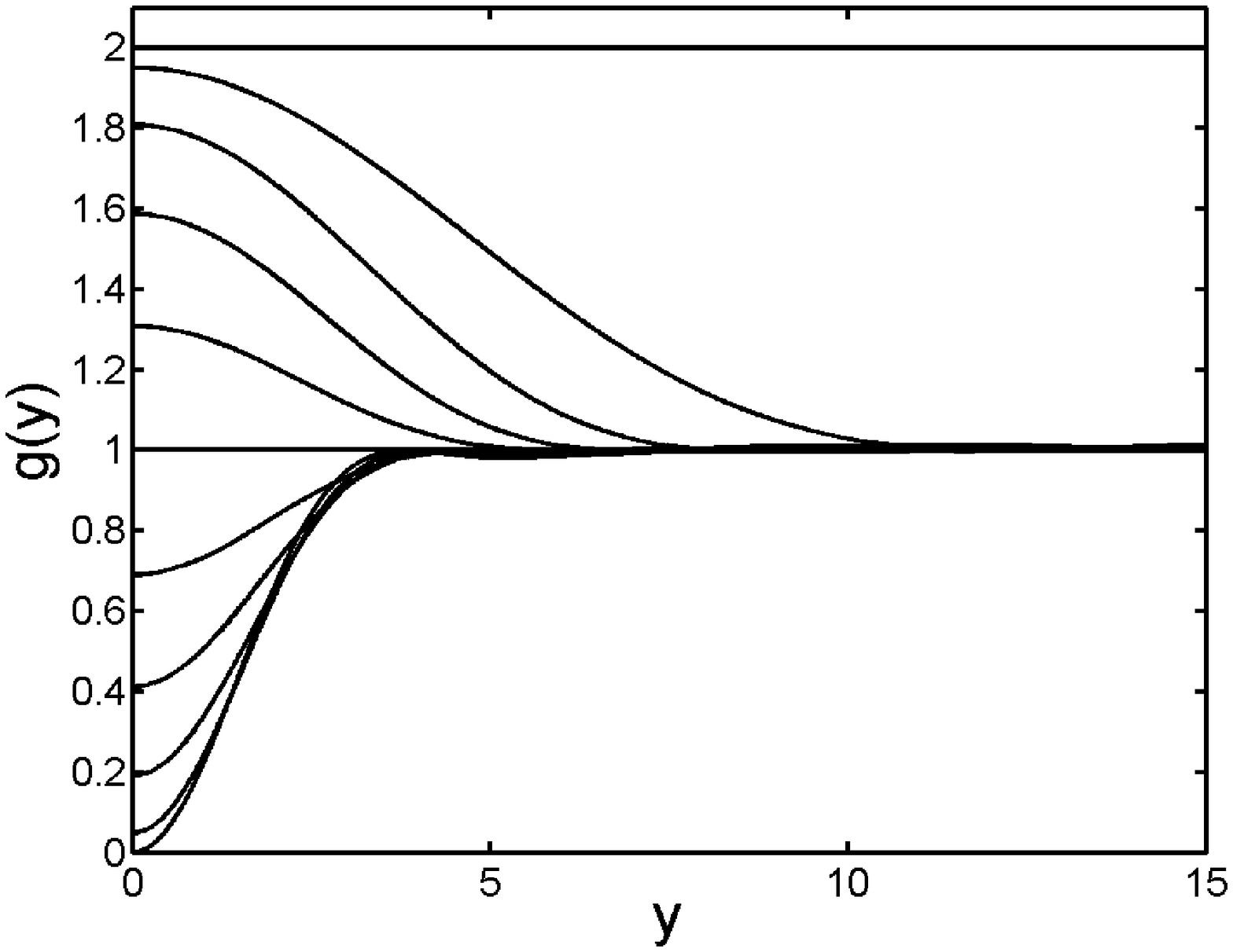}
\end{center}
\end{minipage}
\begin{minipage}[c]{0.49\textwidth}
\begin{center}
\includegraphics[width=\textwidth]{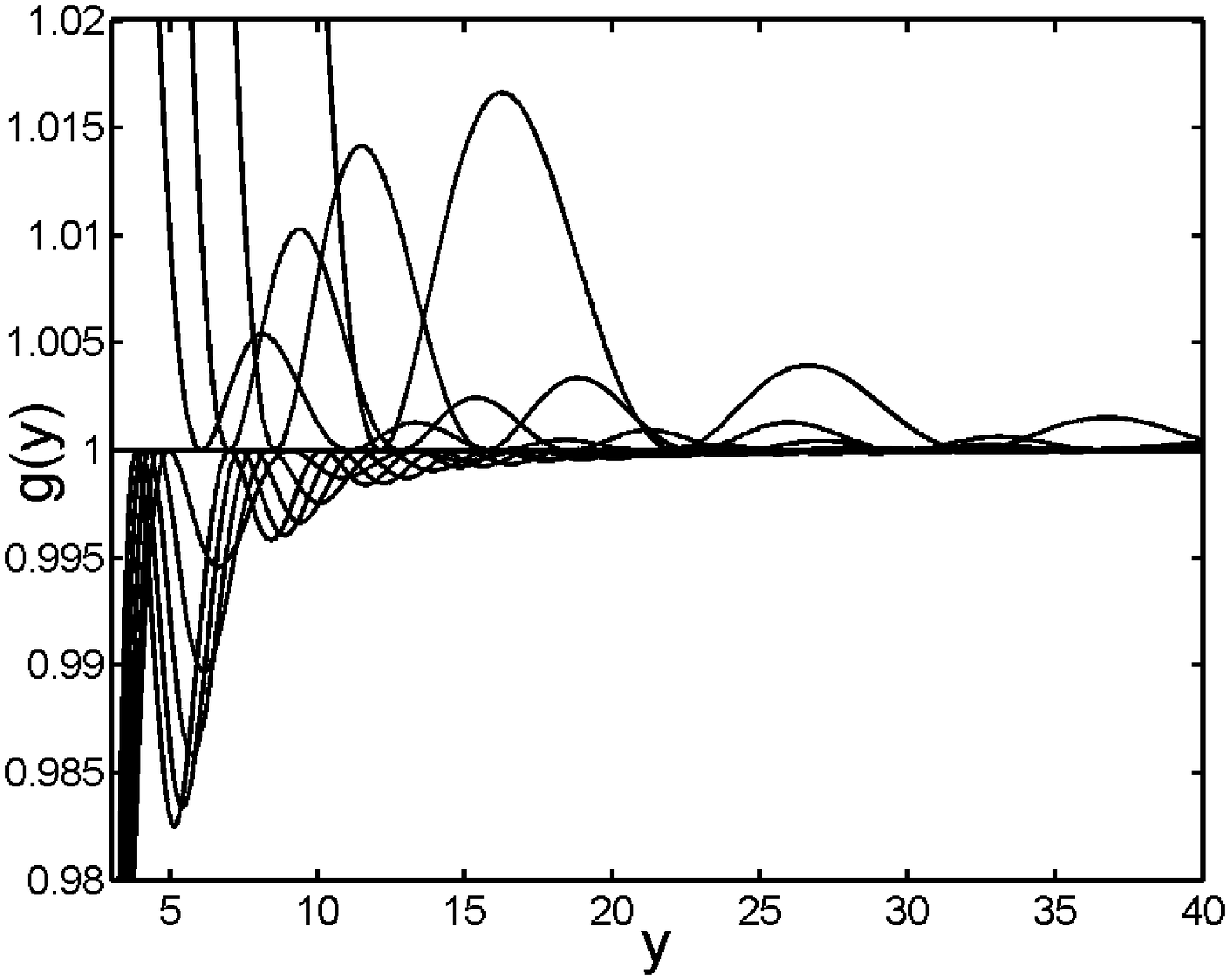}
\end{center}
\end{minipage}
\end{center}
\caption{{\sl (Left)} Pair correlation function $g(y)$ (in scaled units),
Eq.~(\ref{eq:g_zero2}), in $d=2$ for $\alpha=0-1$ ($\alpha$ increases from top to
bottom). {\sl (Right)} The figure shows Friedel-like oscillations of the pair
correlation function, which are present in each case with $\alpha \neq
0,\frac{1}{2}$.}
\label{fig:tzero}
\end{figure}

Figure~\ref{fig:tzero}, on the left hand side, shows the pair correlation
function $g(y)$ at $T=0$ in $d=2$ for $\alpha=0-1$. In the fermionic regime
($\frac{1}{2} <\alpha\leq 1$), the pair correlation function exhibits a
\emph{hole} centred at $y=0$, whose depth decreases with decreasing  $\alpha$
until it disappears at $\alpha=\frac{1}{2}$. For $0<\alpha<\frac{1}{2}$ $g(y)$
shows a \emph{hill} centred at $y=0$, whose height decreases with increasing
$\alpha$. In the bosonic limit ($\alpha=0$), the pair correlation function
$g(y)$ is a constant, $g(y)=2$, which is a known result. Figure \ref{fig:tzero},
on the right hand side, allows us to appreciate the damped Friedel-like
oscillations of the pair correlation function for each non-zero value of the
statistical parameter $\alpha$, except for the case $\alpha=\frac{1}{2}$. 

Making use of Eq.~(\ref{eq:g_zero}) and of the asymptotic properties of the Bessel function $J_\nu(z)$, in the limit $z\gg1$, the wavelength $R$ of the Friedel-like oscillations in the pair correlation function can be derived as:
\begin{equation} \label{eq:periodo_zero}
R=\frac{2\pi}{k_{f,\alpha}}=\sqrt{\pi} \left(\alpha \Gamma\left(d/2\right) \frac{N}{V}\right)^{-\frac{1}{d}}.
\end{equation}
For each dimensionality $d$, at $T=0$ and at fixed density $N/V$, the wavelength $R$ depends explicitly on the statistical parameter $\alpha$. Indeed, $R$ increases with decreasing $\alpha$, and in the bosonic limit ($\alpha=0$) the wavelength becomes infinity ($R=\infty$).

On the basis of Eq.~(\ref{eq:periodo_zero}), one would find a finite wavelength
$R$ also for $\alpha=\frac{1}{2}$. Therefore, the absence of Friedel-like
oscillations in the pair correlation function (cf. Fig.~\ref{fig:tzero}, right
panel, for $T=0$, and Ref.~\cite{Pellegrino:07} for finite temperature) can only
be related to the presence of the vanishing factor $\cos(\alpha\pi)$ in the
original definition of the pair correlation function, Eq.~(\ref{eq:gr}). Indeed,
in this case the factor $\cos(\alpha\pi)$ cancels every deviation of
$g(\textbf{r})$ from unity. In other words, the absence of Friedel-like
oscillations in the pair correlation function in the so-called semionic case
($\alpha=\frac{1}{2}$) is not a consequence of the exclusion statistics in
momentum space, but of the exchange properties implied in our \emph{ad hoc}
generalization of the commutation relations for anyons obeying exclusion
statistics \cite{Pellegrino:07}.  Qualitatively similar results (not shown here)
are obtained in $d=1$ and $d=3$.

Generally, Friedel oscillations in the many-body properties of fermion
assemblies arise from the presence of a discontinuity in the Fermi distribution
function at the Fermi level. This behaviour is characteristic of Wu's
distribution function when $\alpha$ is non-zero. In the literature, it is known
that at low but finite temperature the Friedel oscillations in the charge
density around a screened impurity are exponentially damped with a
characteristic decay length $\lambda^2 k_f$, where $\lambda=\hbar \sqrt{2\pi/(m
k_B T)}$ is the thermal wavelength.  Since at low temperature the behaviour of
Wu's distribution function for non-zero $\alpha$ is equivalent to the
Fermi-Dirac distribution, we find that the Friedel oscillations are exponentially
damped with a characteristic decay length $\lambda^2 k_{f,\alpha}$.

Making use of these remarks, we can derive an approximate form of the pair
correlation function at finite temperature. The idea consists in weakly
modifying the function $g(r)$, Eq.~(\ref{eq:galpha}), which is exact at $T=0$.
We multiply it by a damping factor which has a characteristic decay length:
\begin{equation} \label{eq:decay_l}
L=\lambda^2 k_{f,\alpha}.
\end{equation}
In this way the approximate correlation function can be defined as:
\begin{equation} \label{eq:g_approx}
g(r) = 1+\exp \left( {-\frac{r}{L}} \right) \cos(\alpha \pi) \Gamma\left(\frac{d}{2}\right) \left( \frac{2}{k_{f,\alpha} r}\right)^d J^2_{d/2}(k_{f,\alpha} r).
\end{equation} 

Analogously, at $T=0$ we can rewrite the pair correlation function by means of a change of variable, and we obtain:
\begin{equation} \label{eq:g_zero2approx}
g(y) = 1+\exp \left( {-\frac{y}{Y}} \right)\cos(\alpha \pi) \Gamma\left(\frac{d}{2}\right)  \frac{2^d}{\alpha y^d}   J^2_{d/2}(\alpha^{1/d}y),
\end{equation} 
where the dimensionless quantity $Y$ acts as a characteristic decay length, and is defined as:
\begin{equation} \label{eq:decay_Y}
Y=\alpha^{\frac{1}{d}} \lambda^2 k_f^2. 
\end{equation} 

\begin{figure}[t]
\begin{center}
\begin{minipage}[c]{0.49\textwidth}
\begin{center}
\includegraphics[width=\textwidth]{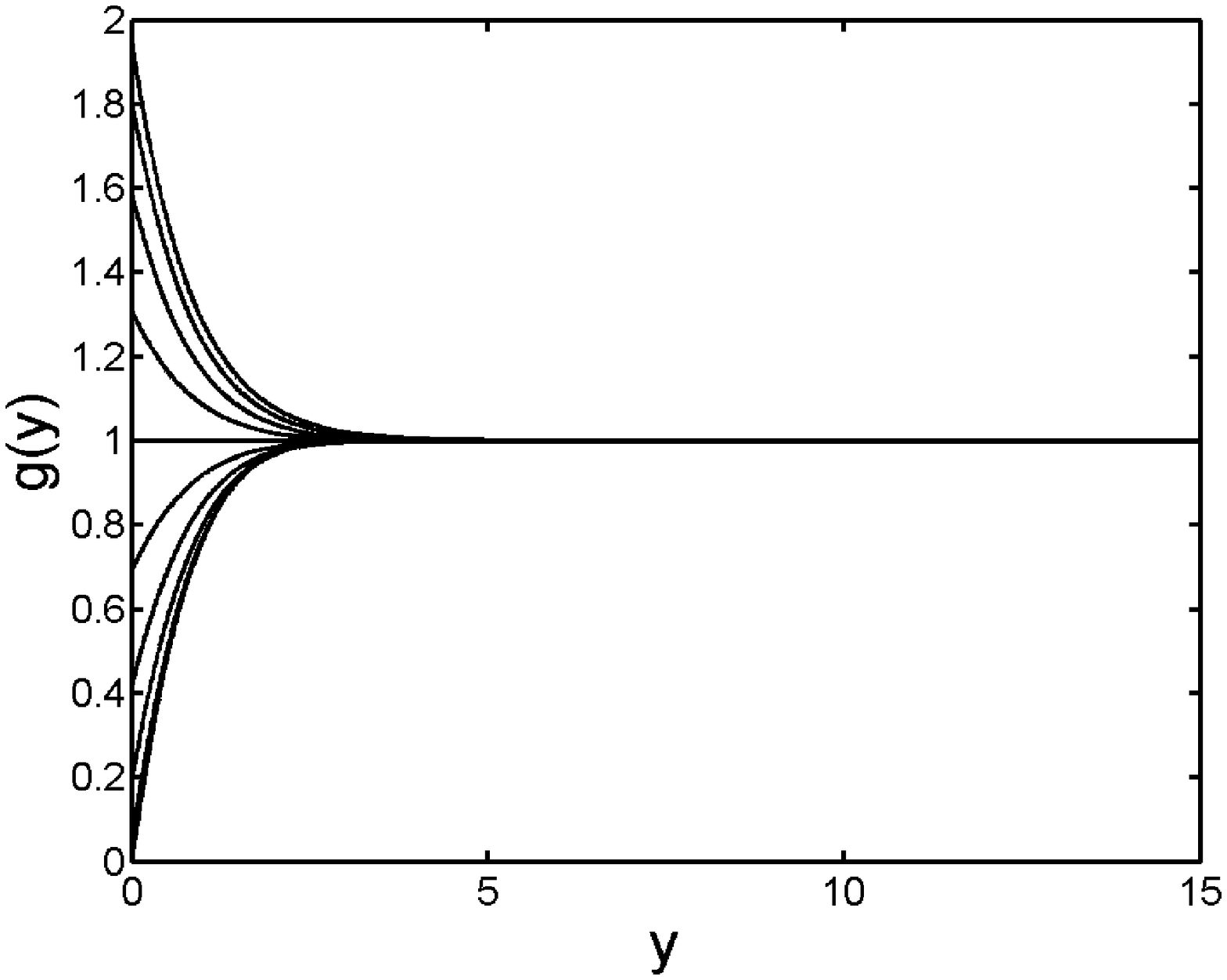}
\end{center}
\end{minipage}
\begin{minipage}[c]{0.49\textwidth}
\begin{center}
\includegraphics[width=\textwidth]{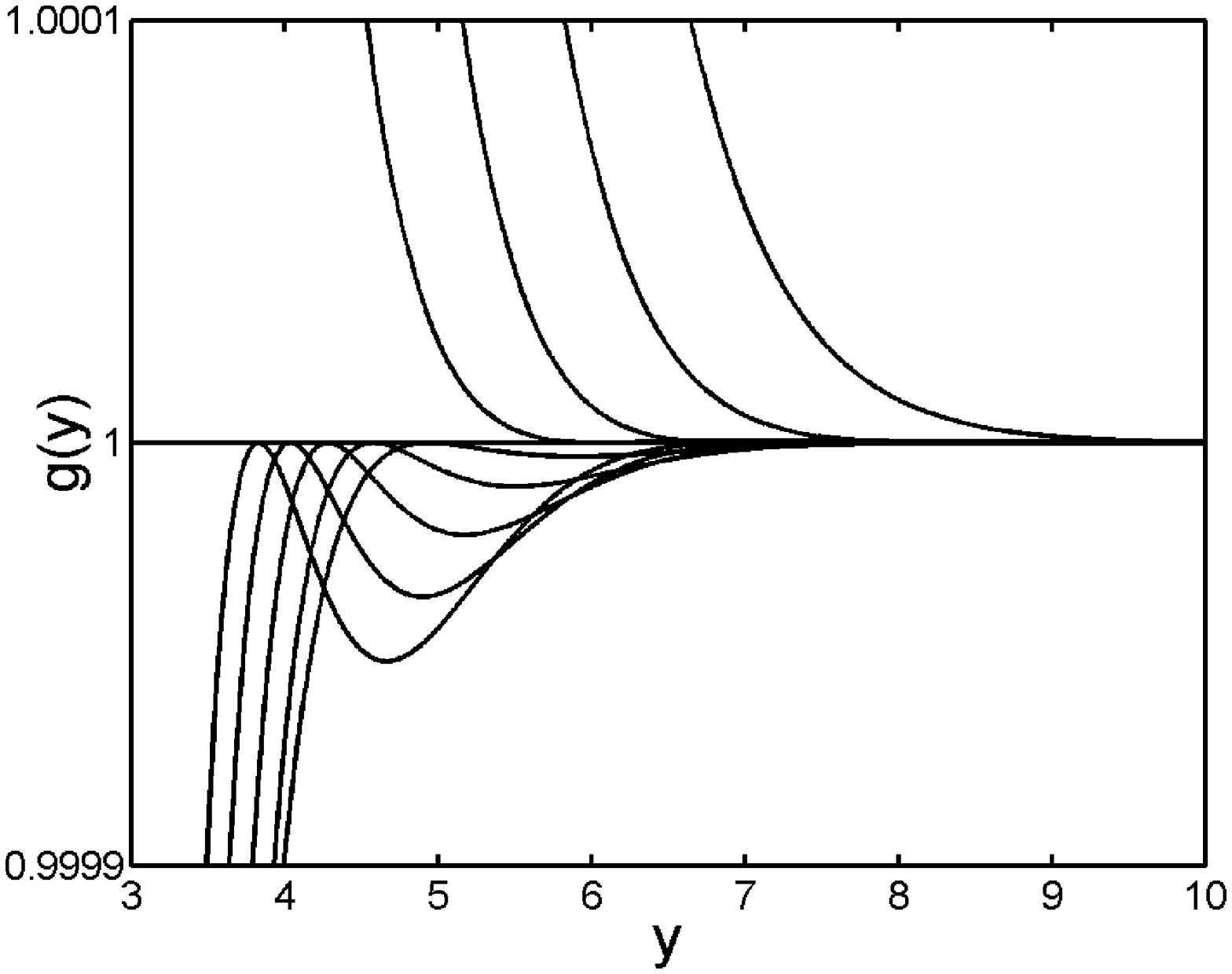}
\end{center}
\end{minipage}
\end{center}
\caption{{\sl (Left)} Approximate pair correlation function $g(y)$ (in scaled
units), Eq. (\ref{eq:g_zero2}), in $d=2$ and with $\lambda^d N/V=1/15$, for $0<
\alpha \leq 1$ ($\alpha$ increases from top to bottom). {\sl (Right)} Showing
Friedel-like oscillations in $g(y)$.}  
\label{fig:approx}
\end{figure}

Figure~\ref{fig:approx} shows the approximate pair correlation function in $d=2$
for $\lambda^d N/V=1/15$. This value of the scaled density is still far from the
classical limit, since the average distance between two particles is comparable
to the thermal wavelength. Likewise, the approximate pair correlation function
at $T=0$ shows a hill centred at $y=0$ when the statistical parameter $\alpha$
is close to the bosonic limit, $0\leq\alpha<\frac{1}{2}$, and shows instead a
hole near the fermionic limit, $\frac{1}{2}<\alpha\leq1$.

Moroever, Figure~\ref{fig:approx} shows that the oscillations are absent only
for $\alpha=0$ and $\frac{1}{2}$. For the cases which are close to the bosonic
limit ($0\leq\alpha<\frac{1}{2}$) the damping term extinguishes completely the
Friedel-like oscillations. On the other hand, for the cases which are close to
the fermionic limit ($\frac{1}{2}<\alpha\leq1$), under the same conditions, the
oscillations suffer from a heavy attenuation but are still visible. It is clear
from Eq.~(\ref{eq:decay_Y}) that under the same conditions (\emph{i.e.}
temperature, density, dimensionality), the damping of the Friedel-like
oscillations depends explicitly on the parameter $\alpha$. Indeed, the value of
the scaling variable $Y$ increases with increasing $\alpha$. Near the bosonic
limit, the characteristic decay length is so short that the oscillations are
completely damped out, whereas close to the fermionic limit the characteristic
decay length is so large that the first Friedel-like oscillations survive.
Pictorially, we can state that the ideal anyonic liquids losing Friedel-like
oscillations in their pair correlation function at low $T$, are those which have
smaller volume within the generalized Fermi surface.

Summarizing, following a proposed generalization of the pair correlation
function $g(\textbf{r})$ for a gas of non-interacting anyons obeying exclusion
fractional statistics in arbitrary dimensions, we have studied the asymptotic
statistical correlations between anyons at $T=0$ and in the limit of low
temperature. We find Friedel-like oscillations in $g(\textbf{r})$ in all
quasi-fermionic cases at $T=0$, except for the value $\alpha=\frac{1}{2}$ of the
statistical parameter. At low temperature, such oscillations are damped
exponentially, with a characteristic decay length which increases with
decreasing density and increasing dimensionality.

\subsection*{Acknowledgements}

This paper was brought to completion during a visit of GGNA and NHM to the
Department of Physics, University of Antwerp. Thanks are due to Professors C.
Van~Alsenoy and D. Lamoen for their warm hospitality. NHM was partially
supported by FWO-Vlaanderen under Project No. G.0425.05.

\begin{small}
\bibliographystyle{pcl}
\bibliography{a,b,c,d,e,f,g,h,i,j,k,l,m,n,o,p,q,r,s,t,u,v,w,x,y,z,zzproceedings,Angilella}
\end{small}

\end{document}